\documentstyle[12pt,epsfig]{article}
\newcommand{\be}{\begin{equation}}
\newcommand{\ee}{\end{equation}}
\newcommand{\bea}{\begin{eqnarray}}
\newcommand{\eea}{\end{eqnarray}}

\def\fun#1#2{\lower3.6pt\vbox{\baselineskip0pt\lineskip.9pt
\ialign{$\mathsurround=0pt#1\hfil##\hfil$\crcr#2\crcr\sim\crcr}}}

\begin{document}
\title{Systematics of $q\bar q$ states, scalar mesons and glueball}
\author{V.V. Anisovich \thanks {Talk given on "HADRON 2001"} \\
{\it St.Petersburg Nuclear Physics Institute}}
\date{}

\maketitle

\begin{abstract}
Basing on the latest results of the PNPI (Gatchina) and QM\&W College
(London) groups, I discuss  systematics of the $IJ^{PC}$ $q\bar q$
states in terms of trajectories on the $(n,M^2)$
plane, where $n$ is the radial quantum number
and $M$ is its mass. In the scalar sector,
which is the most interesting because of the presence of extra  states
with respect to
the $q\bar q$ systematics, I discuss:  1) the results of the
$K$-matrix analysis of the spectra $\pi\pi$, $\pi\pi\pi\pi$, $K\bar K$,
$\eta\eta$, $\eta\eta'$, $\pi\eta$
and characteristics of the resonances in the scalar sector,
2) $q\bar q$-nonet classification of scalar bare states, 3)
accumulation of widths of the $q\bar q$ states by the glueball due to
the overlapping of  $f_0$-resonances at 1200--1700 MeV,
4) systematics of scalar $q\bar q$ states, both
bare states and resonances, on the $(n,M^2)$-plots,
5) constraints on the quark-gluonium content of the resonances
$f_0(980)$, $f_0(1300)$, $f_0(1500)$, $f_0(1750)$,
and the broad state  $f_0(1420^{+150}_{-70})$ from hadronic decays,
6) radiative decays of the $P$-wave $q\bar q$-resonances:
scalars $f_0(980)$, $a_0(980)$, and
tensor mesons $a_2(1320)$, $f_2(1270)$, $f_2(1525)$.
The analysis proves that in the scalar sector we face two exotic
mesons: the light $\sigma$-meson, $f_0(450)$, and the broad state
$f_0(1420^{+150}_{-70})$, which is the descendant of the glueball.
\end{abstract}

{\large {\bf 1. Systematics on the $(n,M^2)$-plots.}}

An important role for the unambiguous interpretation of the data
is played by the $q\bar q$
systematization of the discovered meson states: this may be a
guide for the search for new resonances as well as for establishing
signatures of the existing states.

Here, following \cite{Systematics},
the systematics of $q\bar q$ states is presented in
terms of the $(n,M^2)$ trajectories where $n$ is the radial
quantum number of the $q\bar q$ state and $M$ is its mass.
The trajectories on the $(n,M^2)$
planes are drawn for the $(IJ^{PC})$-states with the positive charge
parity $(C=+)$:  $\pi(10^{-+})$, $\pi_2(12^{-+})$, $\pi_4(14^{-+})$,
$\eta(00^{-+})$, $\eta_2(02^{-+})$, $a_0(10^{++})$, $a_1(11^{++})$,
$a_2(12^{++})$, $a_3(13^{++})$, $a_4(14^{++})$, $f_0(00^{++})$,
$f_2(02^{++})$, and negative one $(C=-)$:  $b_1(11^{+-})$,
$b_3(13^{+-})$, $h_1(01^{+-})$,$\rho(11^{--})$, $\rho_3(13^{--})$,
$\omega /\phi (11^{--})$, $\omega_3  (13^{--})$, see Figs. 1 and 2.
Open points stand for the  predicted states.

The main bulk of information
about the mass region 2000--2400 MeV, which is crucial for drawing the
trajectories, came from the analysis of Crystal Barrel data for
the $p\bar p$ annihilation in flight \cite{new}.

The trajectories on the  $(n,M^2)$-plots
with a good accuracy are linear:
\be
M^2=M_0^2+(n-1)\mu^2.
\ee
$M_0$ is the mass of basic meson and $\mu^2$ is the
trajectory slope parameter: $\mu^2$ is nearly the same for all
trajectories: $\mu^2\simeq 1.2 - 1.3$ GeV$^2$.

Trajectories with the
same $IJ^{PC}$ can be created by the states with different orbital
momenta, with $J=L\pm 1$; in this way they are doubled: these are the
trajectories $(I1^{--})$, $(I2^{++})$, and so on.  Isoscalar states are
formed by  two light flavour components, $n\bar n=(u\bar u+d\bar
d)/\sqrt 2$ and $s\bar s$. Likewise, this also results in doubling
isoscalar trajectories.

The representation of  $(C=-)$-trajectories is thus
determined.  The trajectories are nearly linear with the slope
$\mu^2\simeq 1.3 $ GeV$^2$, with an exception of the $b_J$ sector where
the slope $\mu^2\simeq 1.1 - 1.2 $ GeV$^2$.

For the $C=+$ states, the $\pi_J$-sector is decisively fixed
with the slope $\mu^2\simeq  1.2 $ GeV$^2$. The only state which
breaks linearity of the trajectory is the pion, that is not surprising
because of its specific role in the low-energy physics.

The trajectories in the $\eta_J$-sector are not  unambigously fixed;
in Fig. 1b we show the variant with $\mu^2 =  1.3 $ GeV$^2$.
The uncertainties are mainly due to the region 1700--2000 MeV in the
wave $00^{-+}$: it is the region where one may expect the existence
of pseudoscalar glueball. Indeed, a strong production of the
$00^{-+}$ wave is observed in the radiative $J/\Psi$ decay \cite{Zou}
that may be a  glueball signature (although one should note  that
lattice calculations provide us with a higher value, $\sim 2300$ MeV
\cite{lattice}). For sure, this mass region needs an intensive
study.

The sector of $a_J$ states, $J=0,1,2,3,4$, demonstrates clearly
a set of linear trajectories with
$\mu^2\simeq 1.15 - 1.20 $ GeV$^2$, Figs. 1c, 1e. The same slope is
observed for $f_2$ and $f_4$ mesons, Fig. 1d.

For $f_0$ mesons we have $\mu^2\simeq 1.3 $ GeV$^2$. A superfluous
state for
$q\bar q$-trajectories are the light $\sigma$-meson
\cite{sigma,N/D,D+} and the broad resonance $f_0(1420^{+150}_{-\; 70})$
observed in the K-matrix analysis \cite{YF}: one should consider these
states as candidates for the exotics.

In the recently performed study of the reaction $p\bar p \to
\eta\eta \pi^0\pi^0$ in flight, the resonance with mass
$1880\pm 20$ in the $(12^{-+})$-wave has been declared \cite{pi2}: this
state is also beyond the $\pi_2$-trajectory and should be considered as
a hybrid.

\vspace{0.3cm}

{\large {\bf 2. Scalars.}}

The existence of  superfluous,  with respect to $q\bar q$
systematics, states is a motivation to perform intensive studies on
scalar-isoscalar sector.

{\bf 1) $K$-matrix analysis
and resonances in the scalar-isoscalar sector.}
In the paper \cite{YF},
on the basis of experimental data of GAMS group, Crystal
Barrel Collaboration and BNL group,
the $K$-matrix solution has been found for the waves $00^{++},
10^{++},02^{++},12^{++}$ over the  range
450--1900 MeV. Also the masses and total widths of resonances
have been determined for these waves. The following
states have been seen in the scalar-isoscalar sector:
$f_0(980)$, $f_0(1300)$, $f_0(1500)$, $f_0(1420^{+150}_{-70})$,
$f_0(1750)$. For the scalar-isovector sector,
the analysis \cite{YF} points
to the presence of the following resonances in the spectra:
$ a_0(980)$, $ a_0(1520)$.

The $K$-matrix amplitude takes correctly into account
the threshold singularities of the $00^{++}$
amplitude related to the channels  $\pi\pi$, $\pi\pi\pi\pi$, $K\bar K$,
$\eta\eta$, $\eta\eta'$.
This circumstance
allowed us to reconstruct the
analytical amplitude in the complex-mass region shown in Fig. 3 by
dashed line.
In this
area, with correctly restored analytical structure of
the amplitude $00^{++}$,
we find out the resonance characteristics: the amplitude poles
and decay coupling constants. Besides, we know the $K$-matrix
characteristics such as the $K$-matrix poles.
The $K$-matrix poles are not the amplitude poles, these latter being
connected with
physical resonances, but when the decays are switched off, the
resonance poles turn into the $K$-matrix ones. In the states related to
the $K$-matrix poles there is no cloud of real mesons,
that is due to
the decay processes. This was the reason to call them as "bare states"
\cite{YF}.

Below the mass scale of the $K$-matrix analysis \cite{YF} there is a
pole related to the light $\sigma$-meson (or $f_0(450)$); its position
is shown  in Fig. 3 following the results of the dispersion relation
$N/D$-analysis \cite{N/D} (the mass region validated by this analysis
is also shown in Fig. 3).
Above the mass region of the $K$-matrix analysis
there are resonances $f_0(2030)$, $f_0(2100)$, $f_0(2340)$
\cite{new}.

{\bf 2) Classification of scalar bare states.}
The quark-gluonium
systematics of scalar particles, in terms of bare states, has been
suggested in \cite{Al-An}.  A bare state being a member of
the $q\bar q$ nonet imposes rigid restrictions upon the $K$-matrix
parameters.  The $q\bar q$ nonet of scalars consists of two
scalar-isoscalar states, $f_0^{bare}(1)$ and $f_0^{bare}(2)$,
scalar-isovector meson $a_0^{bare}$ and
scalar kaon $K^{bare}_0$. In the leading order of the
$1/N $-expansion the decays of these four states
into two pseudoscalars
are determined by three parameters only, which are the common constant
$g$, suppression parameter $\lambda$ for strange quark production
(in the limit of a precise $SU(3)_{flavour}$ symmetry
$\lambda=1$) and mixing angle $\varphi$
for the $n\bar n=(u\bar u+d\bar d)/\sqrt{2}$
and $s\bar s$ components in $f_0^{bare}$:
$n\bar n \cos \varphi+s\bar s \sin \varphi $.
The mixing angle defines scalar--isoscalar nonet partners
$f_0^{bare}(1)$ and $f_0^{bare}(2)$:
$ \varphi(1)- \varphi(2)=90^\circ\ .$
Restrictions imposed on coupling constants allow one to fix
unambigously  basic scalar nonet \cite{YF,Al-An}:
\be
1^3P_0q\bar q:
f_0^{bare}(720\pm 100),\; a_0^{bare}(960\pm 30),\;
K_0^{bare}(1220^{+\; 50}_{-150}),\;f_0^{bare}(1260\pm 30)\ ,
\label{nonet}
\ee
as well as mixing angle for $f_0^{bare}(720)$ and $f_0^{bare}(1260)$:
$ \varphi(720)=-70^{\circ}\; ^{+5^\circ}_{-10^\circ} $.

To establish the nonet of first radial excitations,  $2^3P_0q\bar
q$, appeared to be a more difficult task. The $K$-matrix analysis
\cite{YF} gives us
two scalar-isoscalar states at 1200--1650
MeV, $f_0^{bare}(1230^{+150}_{-\; 30}) $ and $f_0^{bare}(1600\pm
50) $; the decay couplings for both of them satisfy  the requirements
imposed for the glueball.
  To resolve this dilemma, we have performed the
systematization of the $q\bar q$
states on the $(n,M^2)$ plot.
Such a systematization
definitely proves that $f_0^{bare}(1600\pm 50) $ is an extra state
for the $q\bar q$ trajectory. In this way,
$f_0^{bare}(1230^{+150}_{-\; 30})$ and $ f_0^{bare}(1810\pm 30)$
must be the $q\bar q$ states.

Then the nonet $2^3P_0q\bar q$ looks as follows:
\be
2^3P_0q\bar q:
f_0^{bare}(1230^{+150}_{-\; 30}),\; f_0^{bare}(1810\pm 30),\;
a_0^{bare}(1650\pm 50),\;K_0^{bare}(1885^{+\; 50}_{-100})\ .
\label{radnonet}
\ee
The decay couplings of $f_0^{bare}(1600)$
to channels
$\pi\pi,K\bar K, \eta\eta, \eta\eta'$ obey the requirements for
the glueball decay. This gives us the reason to consider this state as
the lightest scalar glueball:
\be
0^{++}\; glueball: \qquad
f_0^{bare}(1600\pm 50)\ .
\ee
The lattice calculations are in
reasonable agreement with such a value of the lightest glueball mass.

After the onset of decay channels, the bare states
have transformed into
real resonances. For scalar--isoscalar sector
we observe the following transitions after switching-on the decay
channels:
$f_0^{bare}(720)\pm 100) \to \; f_0(980)$,
$f_0^{bare}(1260\pm 30) \to \;f_0(1300) $,
$f_0^{bare}(1230^{+150}_{-\; 30}) \to \;f_0(1500)$,
$f_0^{bare}(1600\pm 50) \to \;f_0(1420^{+150}_{-70})$,
$f_0^{bare}(1810\pm 30) \to \;f_0(1750)$.

The evolution of bare states into real resonances is illustrated by
Fig. 4: the shifts of amplitude poles on the complex-$M$
plane correspond to  a gradual onset of the decay channels. Technically
it is done by replacing  the phase space $\rho_a$ for
$a=\pi\pi,\pi\pi\pi\pi,K\bar K, \eta\eta, \eta\eta'$ in the
$K$-matrix amplitude
as follows: $\rho_a\to \xi \rho_a $,  the parameter
$\xi$ running in the interval $0\le \xi \le 1$. At $\xi \to 0$ one has
pure bare states, while the limit $\xi \to 1$ gives us the position of
real resonance.

Note that the broad state is denoted in \cite{YF} as
$f_0(1530^{+\;90}_{-250})$
that is the averaged value for three solutions found in \cite{YF};
the value of the mass given in Figs. 3 and 4,
$M=(1420^{+\;150}_{-70})-i(540\pm 80)$ MeV, corresponds to the solution
for which the scalar glueball is located near 1600 MeV.

{\bf 3) The overlapping of $f_0$-resonances at
1200--1700 MeV: accumulation of widths of  $q\bar q$ states by the
glueball. }
The appearance of broad resonance is not at all an occasional
phenomenon. It has originated as a result of a mixing of states which
are due to the decay processes, namely,  transitions
$f_0(m_1)\to real\; mesons\;
\to f_0(m_2)$. These transitions
result in a specific phenomenon, that is, when several resonances
overlap, one of them accumulates the widths of neighbouring resonances
and transforms into a broad state.

This phenomenon has been observed in
\cite{YF} for the scalar-isoscalar states,
and  the following scheme has been suggested in
\cite{glueball}: the broad state $f_0(1420^{+150}_{-70})$
is the descendant of a pure glueball, which being in the neighbourhood
of $q\bar q$ states accumulated their widths and transformed into a
mixture of the gluonium and $q\bar q$ states. In \cite{glueball}
this idea has been applied for four resonances
$f_0(1300)$, $f_0(1500)$, $f_0(1420^{+150}_{-70})$ and $f_0(1750)$,
by using the language of the $q\bar q$ and $gg$ states for consideration
of the decays  $f_0 \to q\bar q, gg$
and mixing processes $f_0(m_1) \to q\bar q, gg\to f_0(m_2)$.
According to \cite{glueball}, the gluonium component is
mainly shared between three resonances,  $f_0(1300)$, $f_0(1500)$,
$f_0(1420^{+150}_{-70})$, so every state is a mixture
of $q\bar q$ and $gg$ components, with roughly equal percentage
of the gluonium (about 30-40\%).

The accumulation of widths of overlapping resonances by one of them is
a well-known effect in nuclear physics.
In meson physics this phenomenon can play an important role, in
particular for exotic states which are beyond the $q\bar q$
systematics. Indeed, being among the $q\bar q$ resonances, the exotic
state creates a group of overlapping resonances.
The exotic state, which is not orthogonal
to its neighbours, after having accumulated the "excess" of width turns
into a broad state. This
broad resonance should be accompanied  by narrow states which are
the descendants of states from which the widths have been taken off. In
this way, the existence of a broad resonance accompanied by narrow
ones may be a signature of exotics. This possibility, in context of
searching for exotic states, has been discussed in
\cite{exotic}.

The broad state may be one of the components which form the
confinement barrier:  the broad states after accumulating the widths of
neighbouring resonances play for these latter the role of locking
states. Evaluation of the mean radii squared of the broad state
$f_0(1420^{+150}_{-70})$ and its neighbours-resonances, performed
in \cite{exotic} on the basis of the GAMS data, argues in
favour of this idea, for the radius of $f_0(1420^{+150}_{-70})$
is significantly larger than that of $f_0(980)$ and $f_0(1300)$
thus making it possible for $f_0(1420^{+150}_{-70})$
to be the locking state.

{\bf 4) Systematics of the $q\bar q$
scalar states on the $(n,M^2)$ plot.}
As is stressed above,
the systematics of $q\bar q$ states on the $(n,M^2)$ plot
argues that the broad state $f_0(1420^{+150}_{-70})$
and its predecessor $f_0^{bare}(1600\pm 50)$ are beyond the
$q\bar q$ classification.
We plot in Fig. 5a
the $(n,M^2)$-trajectories for $f_0$,
$a_0$ and $K_0$ states. All trajectories are roughly linear,
and they clearly represent the
states with dominant $q\bar q$ component.  It is seen that one of the
states, either $f_0(1420^{+150}_{-70})$ or $f_0(1500)$, is superfluous
for the $q\bar q$ systematics. Looking at the $(n , M^2)$-trajectories
of bare states, Fig. 5b, one can see that just $f_0^{bare}(1600)$ does
not fall onto any linear $q\bar q$ trajectory. So it would be
natural to conclude that the state $f_0^{bare}(1600)$ is an exotic one,
i.e. the glueball.

For resonances belonging to linear trajectories  (Fig. 5a) the
$q\bar q$ component is  dominanting. The scalar-isoscalar resonances
$f_0(1300)$,  $f_0(1500)$ contain a considerable gluonium component, and
certain gluonium admixture exists in $f_0(1750)$.
The location of the $f_0(980)$ pole near $K\bar K$ threshold allows one
to suspect the existence of an admixture of the $K\bar K$-component
in this resonance.
To investigate this admixture the precise measurements of the $K\bar
K$ spectra in the interval 1000---1150 MeV are necessary: only these
spectra could shed the light on the role of the long-range $K\bar K$
component in $f_0(980)$.

{\bf 5) Quark-gluonium content of resonances
$f_0(980)$, $f_0(1300)$, $f_0(1500)$, $f_0(1750)$
and the broad state  $f_0(1420^{+150}_{-70})$
from hadronic decays.}
The $K$-matrix analysis does not supply us with coupling constants
of the resonance decay in a direct way. To find them out, additional
calculations are needed to know the residues of amplitude poles
related to resonances. Such calculations have been carried out
in \cite{width}  for the channels
$f_0\to \pi\pi, \pi\pi\pi\pi, K\bar K, \eta\eta,\eta\eta'$.
The conclusion is as follows
\cite{content}: the decays couplings to the channels
$\pi\pi, K\bar K, \eta\eta,\eta\eta'$ do not provide us with a unique
solution for absolute weight of the $n\bar n$, $s\bar s$ and
gluonium components but give us relative weights only.
The mixing angle $\varphi$ which enters the quark wave function
$q\bar q=n\bar n\cos\varphi +s\bar s\sin\varphi $ can be evaluated as a
function of the decay couplings for gluonium and quarkonium components
$G(gg\to hadrons)/g(q\bar q\to hadrons)$. The
ratio of the couplings squared was
conventionally called in \cite{content} as probability for the gluonium
component in the $f_0$-meson: $W\equiv G^2/g^2$. The following
 relations for $\varphi$ verus $W$ have been found \cite{content}:
\bea
 \varphi [f_0(980)]\simeq -67^\circ \pm 57^\circ \sqrt {W(980)},\qquad
 \varphi [f_0(1300)]\simeq -5^\circ \pm 28^\circ \sqrt {W(1300)},
\nonumber \\
 \varphi [f_0(1500)]\simeq 8^\circ \pm 16^\circ \sqrt {W(1500)},\qquad
 \varphi [f_0(1750)]\simeq -27^\circ \pm 42^\circ \sqrt {W(1750)}.
\eea
A large admixture of the gluonium, $W\le 0.4$, may be expected for
$f_0(1300)$, $f_0(1500)$, $f_0(1750)$, but it should be considerably
less in $f_0(980)$, $W(980)\le 0.20$.

The analysis \cite{content} proves that
 $f_0(1420^{+150}_{-70})$    contains
the $q\bar q $ in the flavour singlet state only:
\be
\varphi [f_0(1420^{+150}_{-70})] \simeq 37^\circ\ ,
\ee
 that perfectly agrees with its gluonium origin:
This value of mixing angle practically does not  depend
on the percentage of the $(q\bar q)_{singlet}$ and gluonium components
in the broad state.

{\bf 6) Radiative decays of the $P$-wave $q\bar q$-mesons.}
The investigation of radiative decays is a powerful tool for
establishing the quark structure of hadrons.  At the early stage of the
quark model, the radiative decays of vector mesons provided strong
arguments in favour of the idea of constituent quark, a universal
object for mesons and baryons \cite{Vectors}.
The radiative decays of the $1^3P_J q\bar q$ mesons are
equally important for the verification of the $P$-wave multiplet.

In Ref. \cite{tensor},
 partial widths of the decays $f_0(980)\to \gamma\gamma$
and $a_0(980)\to \gamma\gamma$ have been calculated assuming
$f_0(980)$ and $a_0(980)$ to be dominantly  $q\bar q$ states, that
is, $1^3P_0q\bar q$ mesons. The results of the calculation agree
well with experimental data. On the basis of
experimental data  for the decays
$\phi(1020)\to \gamma f_0(980)$ and $f_0(980)\to \gamma\gamma$
the $n\bar n/s\bar s$ content of $f_0(980)$ has been found. Assuming
the flavour wave function in the form $n\bar n \cos \varphi+s\bar s
\sin \varphi$, the experimental data has been described  with two
possible values of mixing angle: either $\varphi
[f_0(980)]=-48^\circ\pm 6^\circ$ or $\varphi [f_0(980)]=85^\circ\pm
4^\circ$ (negative value is more preferable), see Fig. 6a where the
allowed region of $\varphi$ versus $R^2_{f_0(980)}$ is shown.

The dominance of the quark-antiquark state does not exclude the
existence of other components in $f_0(980)$
on the level $10\% -20\% $, the glueball or
long-range $K\bar K$ component.
The existence of the long-range $K\bar K$ component or that of gluonium
in the $f_0(980)$ results in a decrease of the $s\bar s$ fraction in
the $q\bar q$ component: for example, if the long-range $K\bar K$ (or
gluonium) admixture is of the order of 15\%, the data require either
$\varphi=- 45^\circ \pm 6^\circ$ or $\varphi=83^\circ \pm 4^\circ$.

There is no problem with the  description of the decay $a_0(980)\to
\gamma\gamma$ within the hypothesis about  $q\bar q$ origin of
the  $a_0(980)$: the data are in a good agreement with the results
of the calculation by using $R^2_{a_0(980)} \sim 10$ -- $17$ GeV$^{-2}
$.

Although direct calculations  of widths of radiative
decays agree well with the hypothesis that the $q\bar q$ component
dominates $f_0(980)$ and $a_0(980)$, to determine reliably these mesons
as members of the $1^3P_0q\bar q$ multiplet one more step is necessary.
We have to prove that radiative decays of tensor mesons
$a_2(1320)$, $f_2(1270)$,  $f_2(1525)$ can be calculated within the
same approach and the same technique as it was carried out for
 $f_0(980)$ and $a_0(980)$. Tensor mesons
$a_2(1320)$, $f_2(1270)$,  $f_2(1525)$ are the basic members of the
$P$-wave $q\bar q$ multiplet, and  the existence of  tensor
mesons had been used to suggest quark--antiquark classification for
four $P$-wave nonets \cite{Pqq1}. Under this motivation,
partial widths of the tensor $q\bar q$ states
$a_2(1320)\to \gamma\gamma$, $f_2(1270)\to \gamma\gamma$ and
$f_2(1525)\to \gamma\gamma$ have been calculated \cite{tensor}:
the agreement with data has been reached for all calculated partial
widths, with similar radial wave functions,
that indicates definitely that both scalar ($f_0(980)$,
$a_0(980)$) and tensor ($a_0(1320)$, $f_2(1270)$,  $f_2(1525)$) mesons
belong to the same $P$-wave $q\bar q$ multiplet.  In Fig. 6b one can
see the region of magnitudes $(\varphi_T,R^2_T)$ allowed by  data on
the decays $f_2(1270)\to \gamma\gamma$ and $f_2(1525)\to \gamma\gamma$;
here $\varphi_T$ is mixing angle for
$\psi_{f_2(1270)}=\cos \varphi_T n\bar n+ \sin \varphi_T s\bar s$ and
$\psi_{f_2(1525)}=-\sin \varphi_T n\bar n+ \cos \varphi_T s\bar s$ and
$R_T$ is the tensor-meson radius.

{\bf 7) Exotics in scalar-isoscalar sector.}
The established $q\bar q$ systematics of scalar mesons
in terms of bare states fixes  two nonets:
$1^3P_0q\bar q$ and $2^3P_0q\bar q$.
The resonances which are
the descendants of pure $q\bar q$ states are located on linear
trajectories in the $(n,M^2)$-plane. The $q\bar q$ systematics reveals
two extra states which are the light $\sigma$-meson, with mass
$\sim  450$ MeV, or $f_0(450)$,
and broad state $f_0(1420^{+150}_{-70})$. The
broad state is the descendant of a pure glueball state which
accumulated the widths of neighbouring $q\bar q$ resonances.
The origin of the $\sigma$-meson is questionable.

In the paper \cite{eyewit}, a hypothesis is discussed that the
$\sigma$-meson owes
its origin to strong singlarity in the confinement amplitude: the
large-$r$ behaviour of the confinement scalar potential $V(r) \sim r$
evokes a strong $t$-channel singularity, $1/t^2$. It was assumed in
\cite{eyewit} that the singularity of this kind exists at every
colour state of the confinement $q\bar q$ ladder; then, in the white
state related to the $\pi\pi$ channel, the unitarization of singular
block might reduce the singularity strenth, thus providing the pole
near the $\pi\pi$ threshold, at ${\rm Re}\,s\sim
4\mu_\pi^2 $, that corresponds to the $\sigma$-meson.

The analysis of data on the
decays $D^+ \to \pi^+\pi^+\pi^-$  and $D^+_s \to
\pi^+\pi^+\pi^-$ \cite{D+}
 agrees with such an idea.
The $s\bar s$ component at $f_0(980)$ has been
evaluated in \cite{D-meson}
by comparing the branching ratios $D_s^+ \to\pi^+\phi(1020)$
and $D_s^+ \to\pi^+f_0(980)$  being about  50\%
(this estimate agrees with  hadronic and radiative
decays of $f_0(980)$ for the solution with negative mixing angle
$\varphi$).  The ratio of yields of $f_0(450)$ and $f_0(980)$ in the
reaction  $D^+ \to \pi^+\pi^+\pi^-$ tells us that $f_0(450)$ is
dominantly the $n\bar n$ system. The confinement ladder should be
formed by the light quarks, $(u,d)$, see for example \cite{Gribov}: in
this sense, the structure of $f_0(450)$ is just as it was expected, if
it originates from the confinement ladder, as it was supposed in
\cite{eyewit}.

{\bf Acknowledgement.} Thanks are due to A.V. Anisovich, D.V. Bugg,
L.G. Dakhno, D.I. Melikhov, V.A. Nikonov, A.V. Sarantsev for
numerous discussions. The work is suppored by the RFFI grant N 01-02-17861.


\begin{figure}
\centerline{\epsfig{file=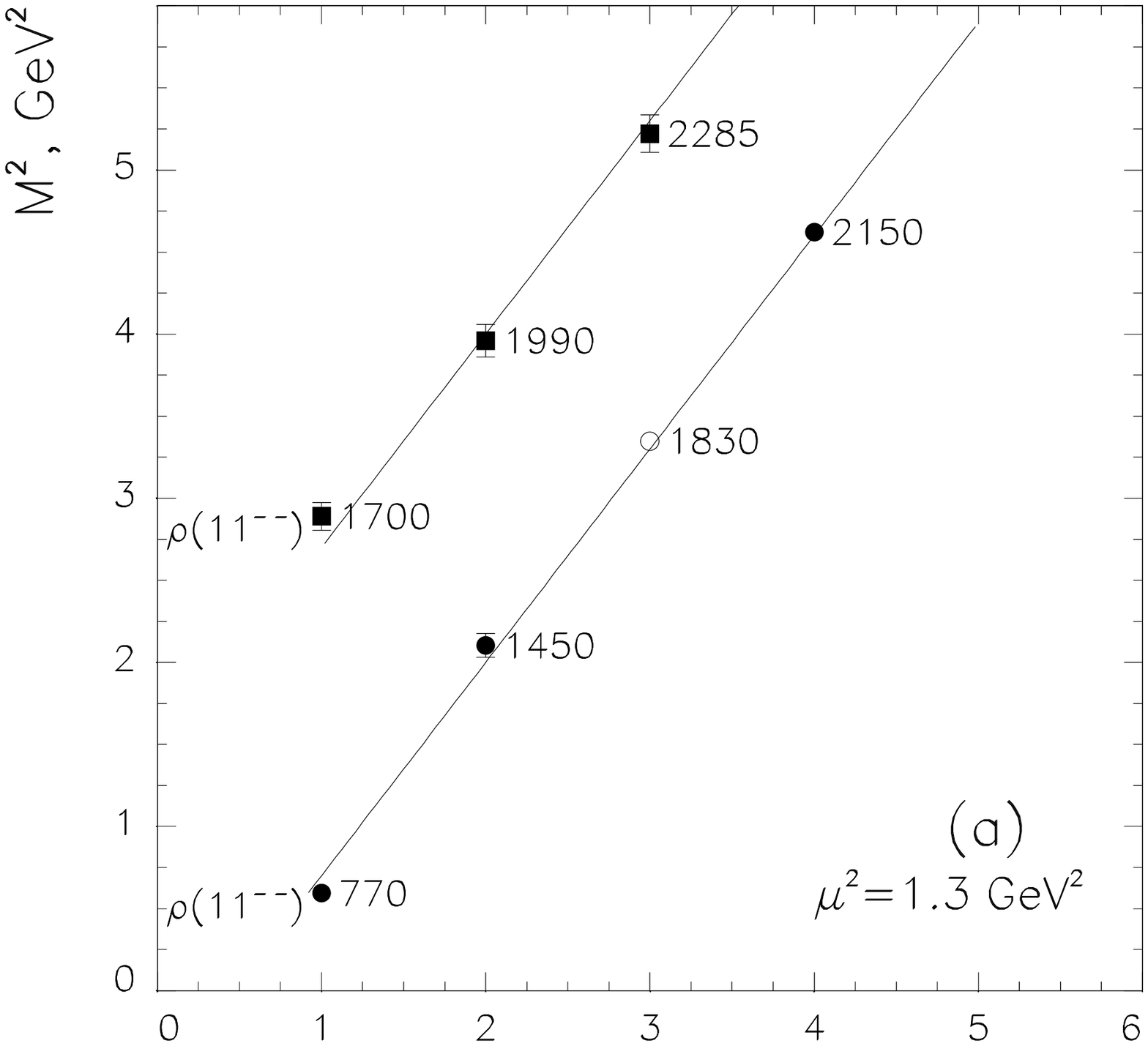,width=8cm}\hspace{-1.5cm}
            \epsfig{file=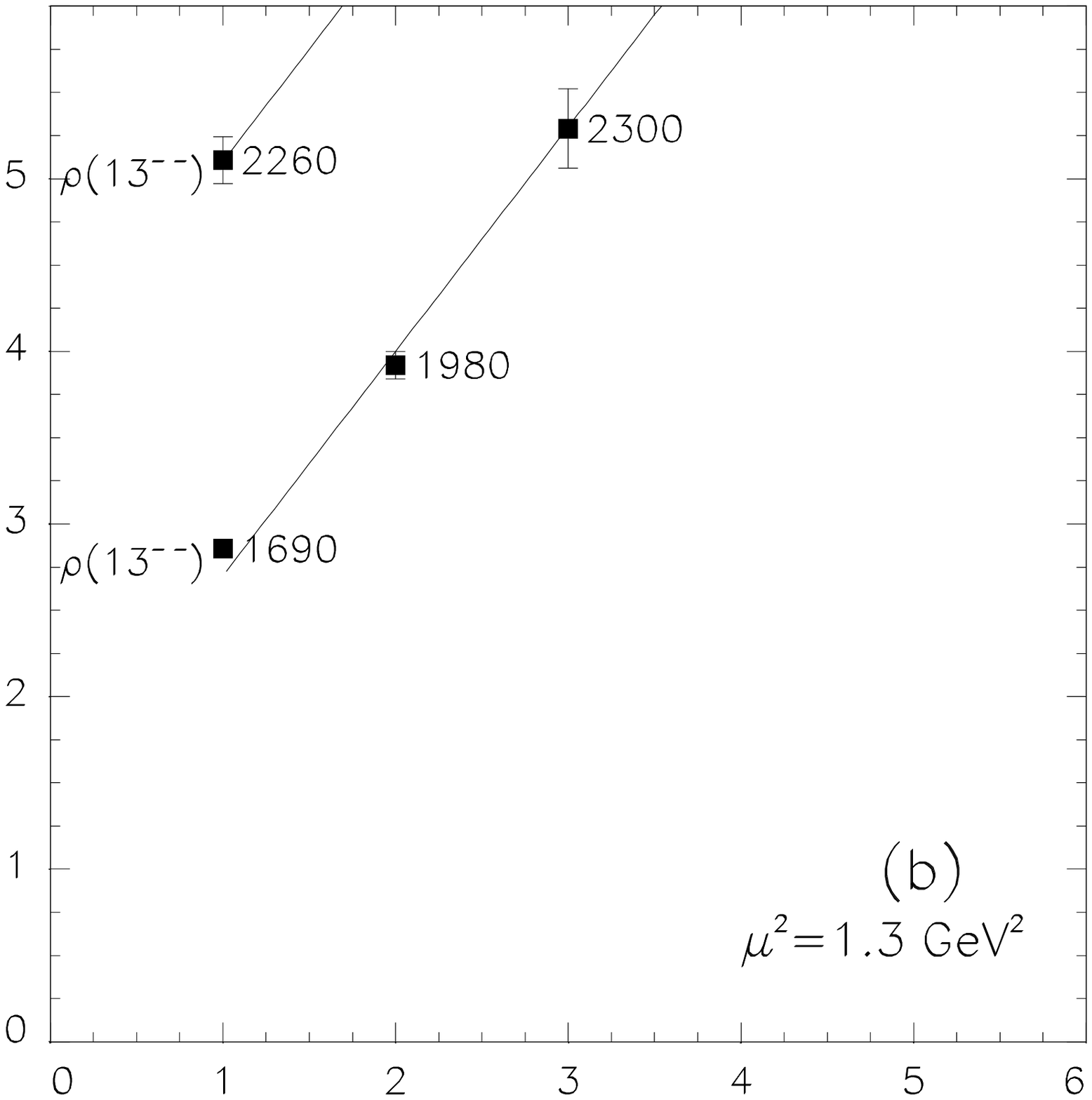,width=8cm}}
\vspace{-1.5cm}
\centerline{\epsfig{file=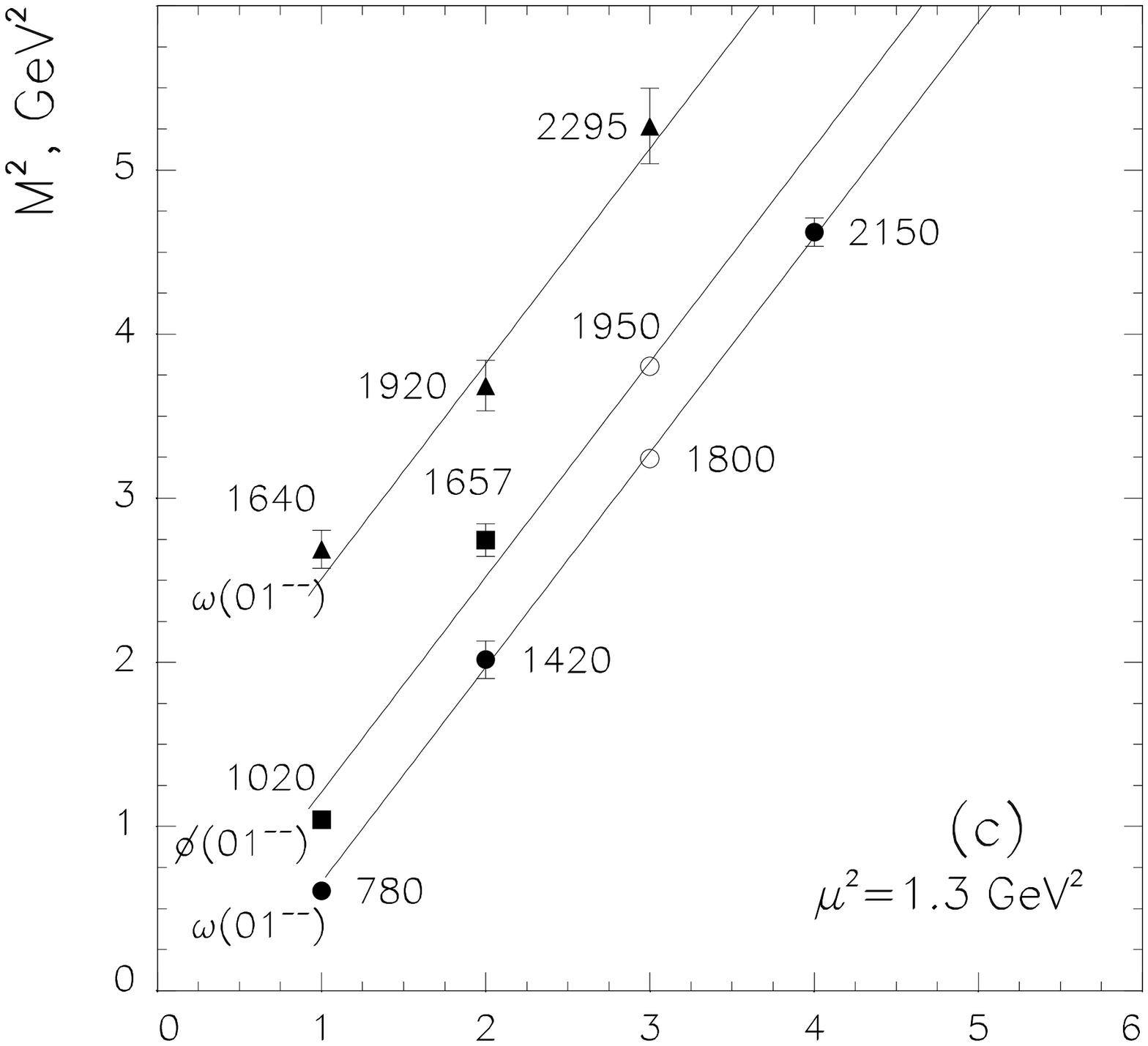,width=8cm}\hspace{-1.5cm}
            \epsfig{file=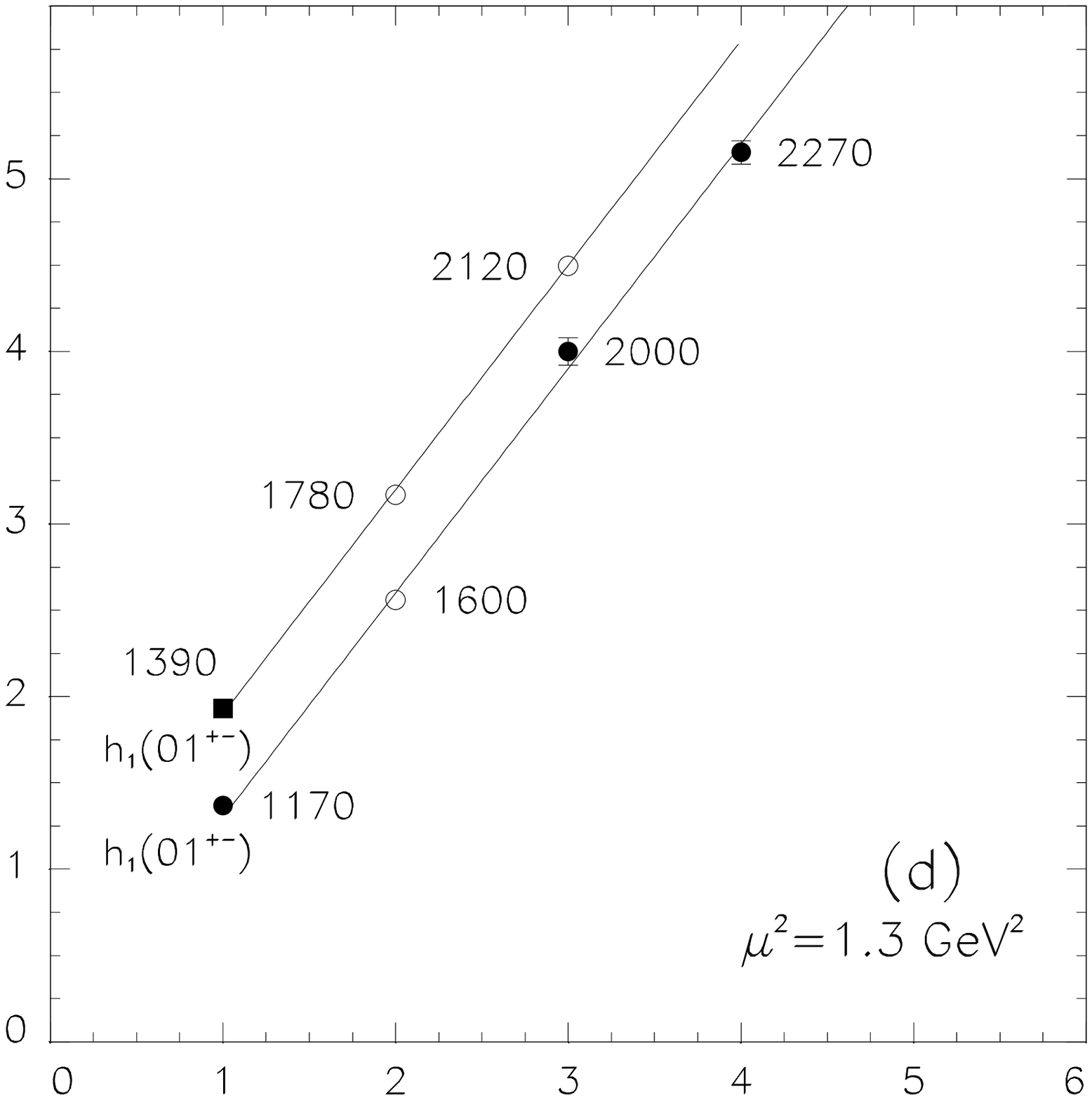,width=8cm}}
\vspace{-1.5cm}
\centerline{\epsfig{file=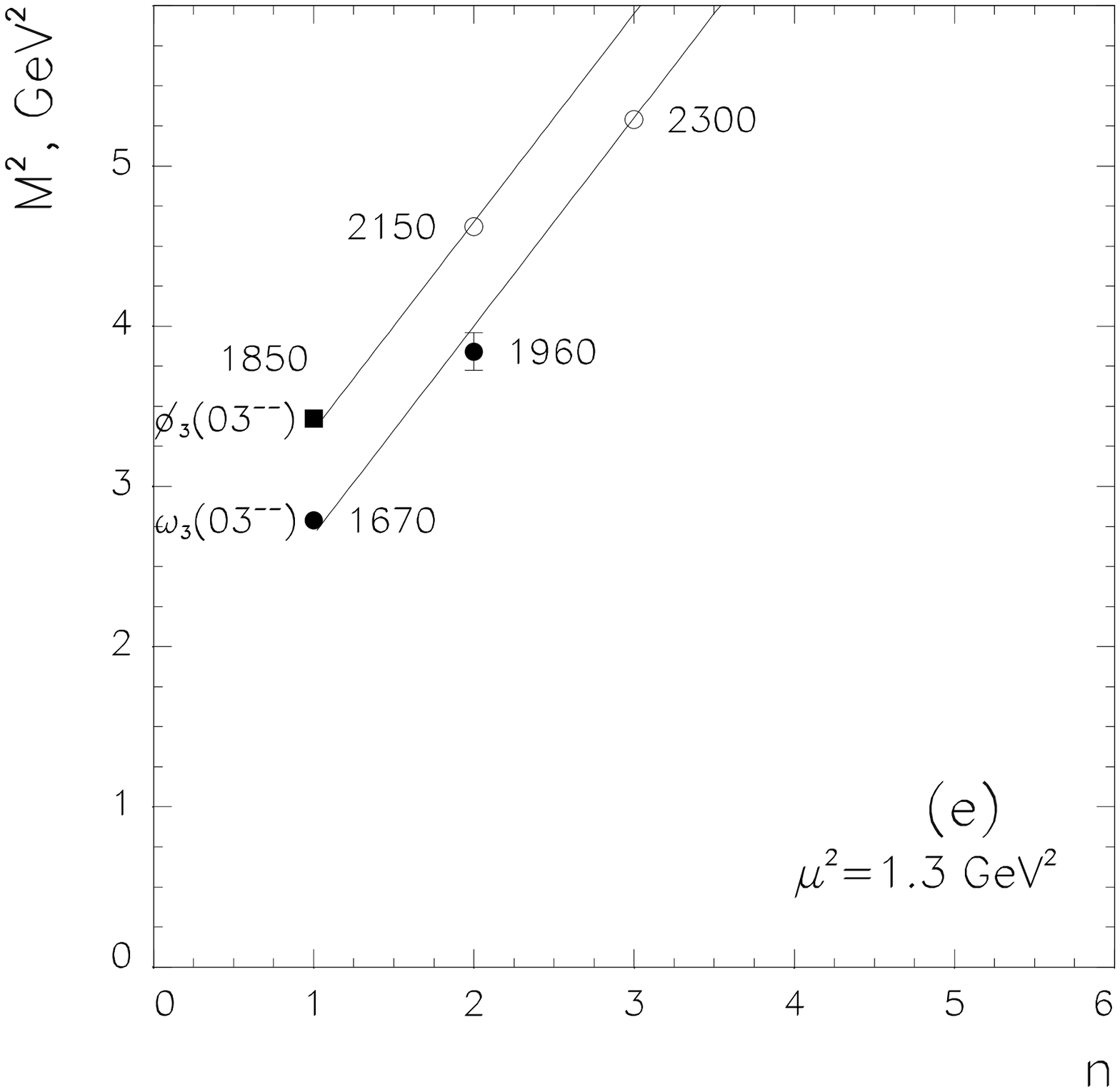,width=8cm}\hspace{-1.5cm}
            \epsfig{file=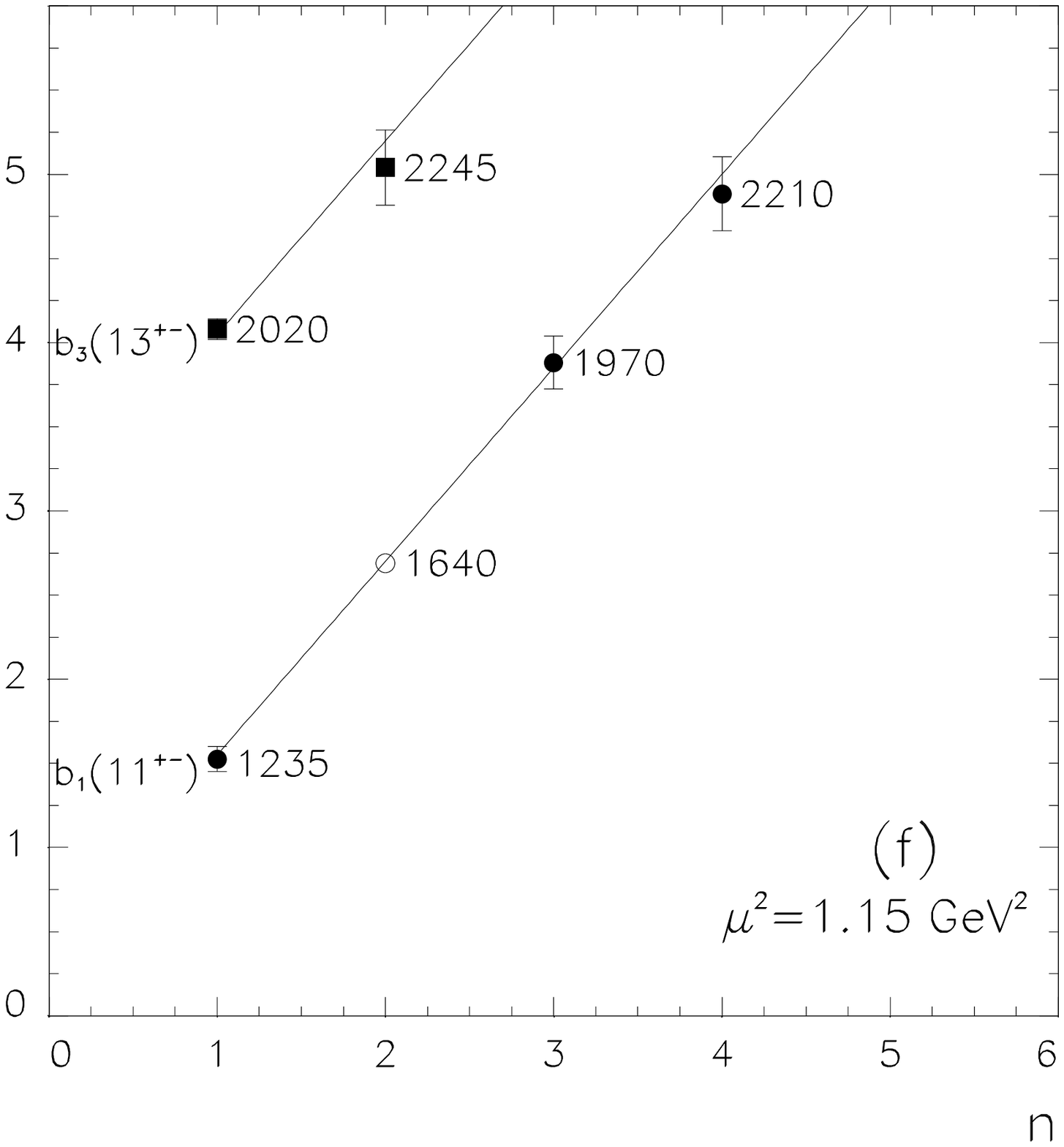,width=8cm}}
\caption{Trajectories of the $(C=-)$-states on the $(n, M^2)$ plane.}
\end{figure}

\newpage

\begin{figure}[h]
\centerline{\epsfig{file=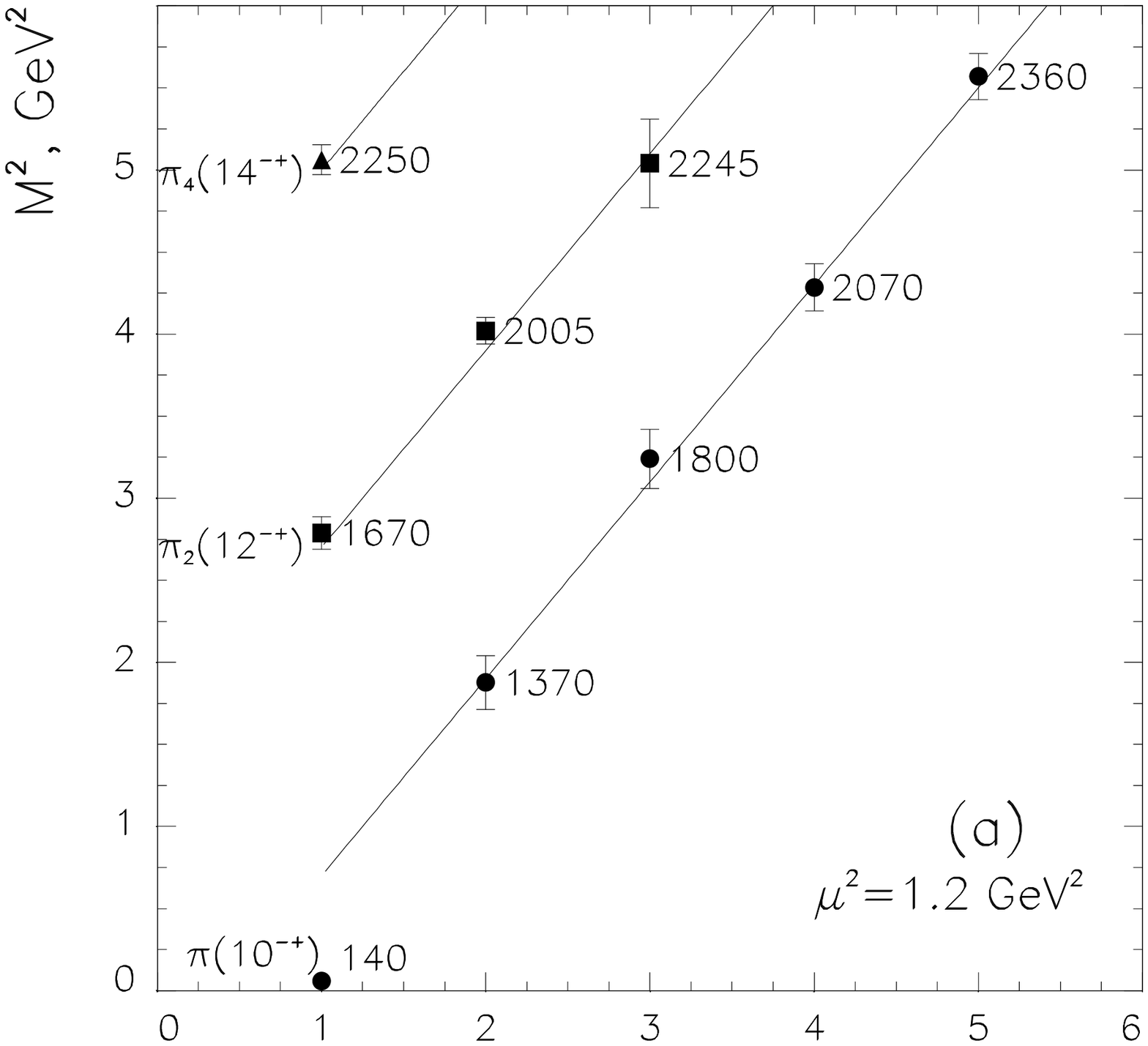,width=8cm}\hspace{-1.5cm}
            \epsfig{file=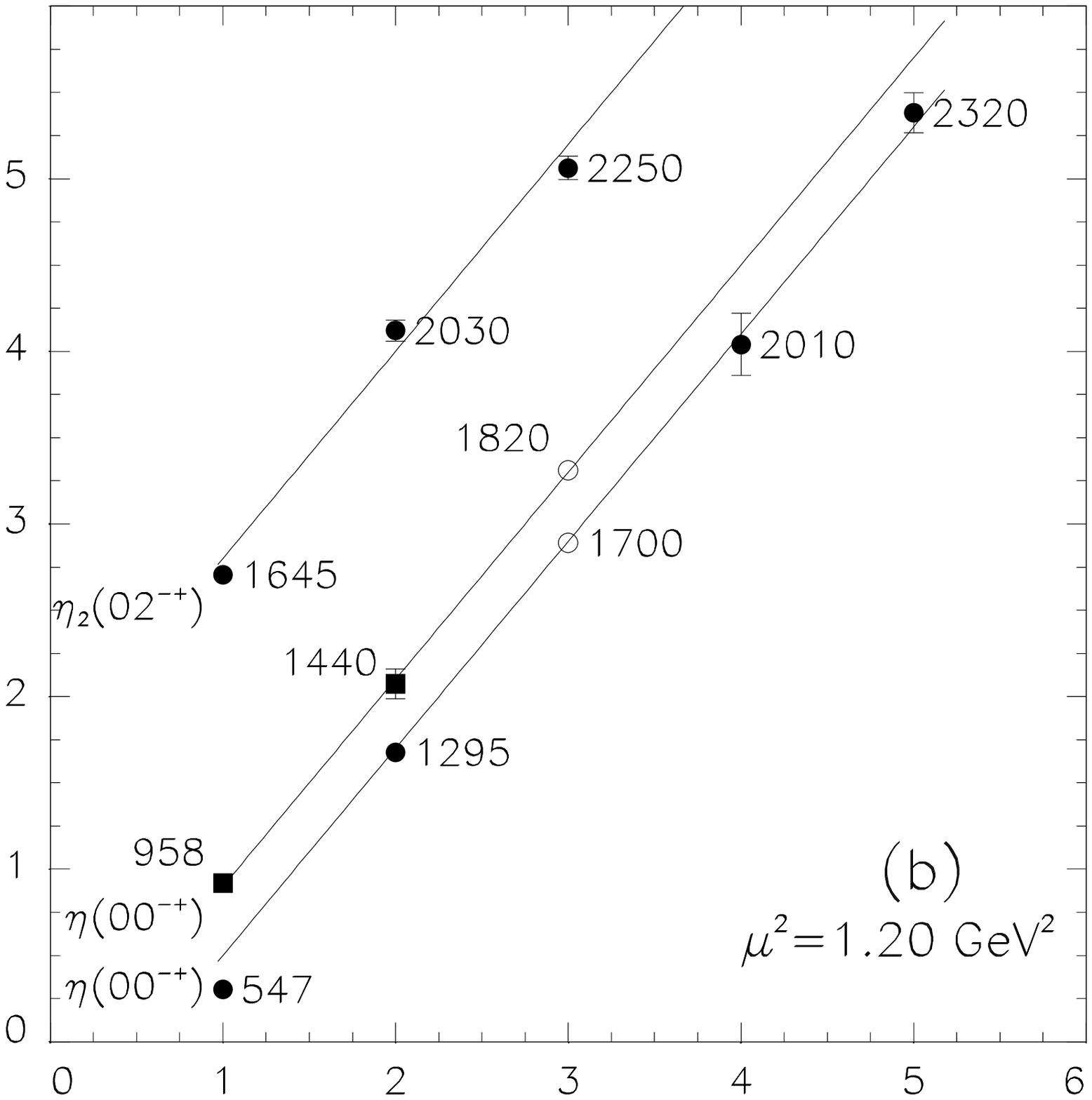,width=8cm}}
\vspace{-1.5cm}
\centerline{\epsfig{file=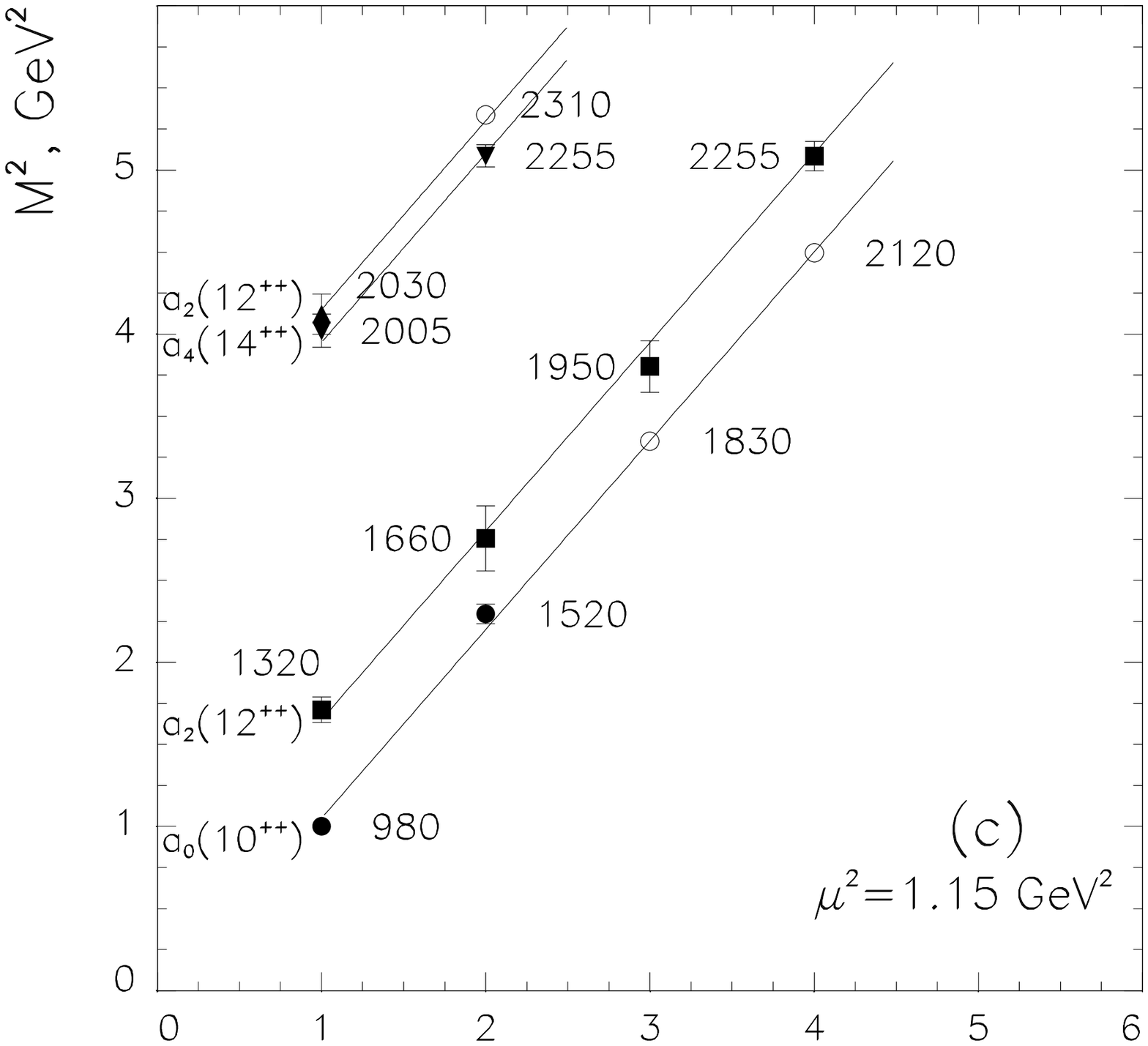,width=8cm}\hspace{-1.5cm}
            \epsfig{file=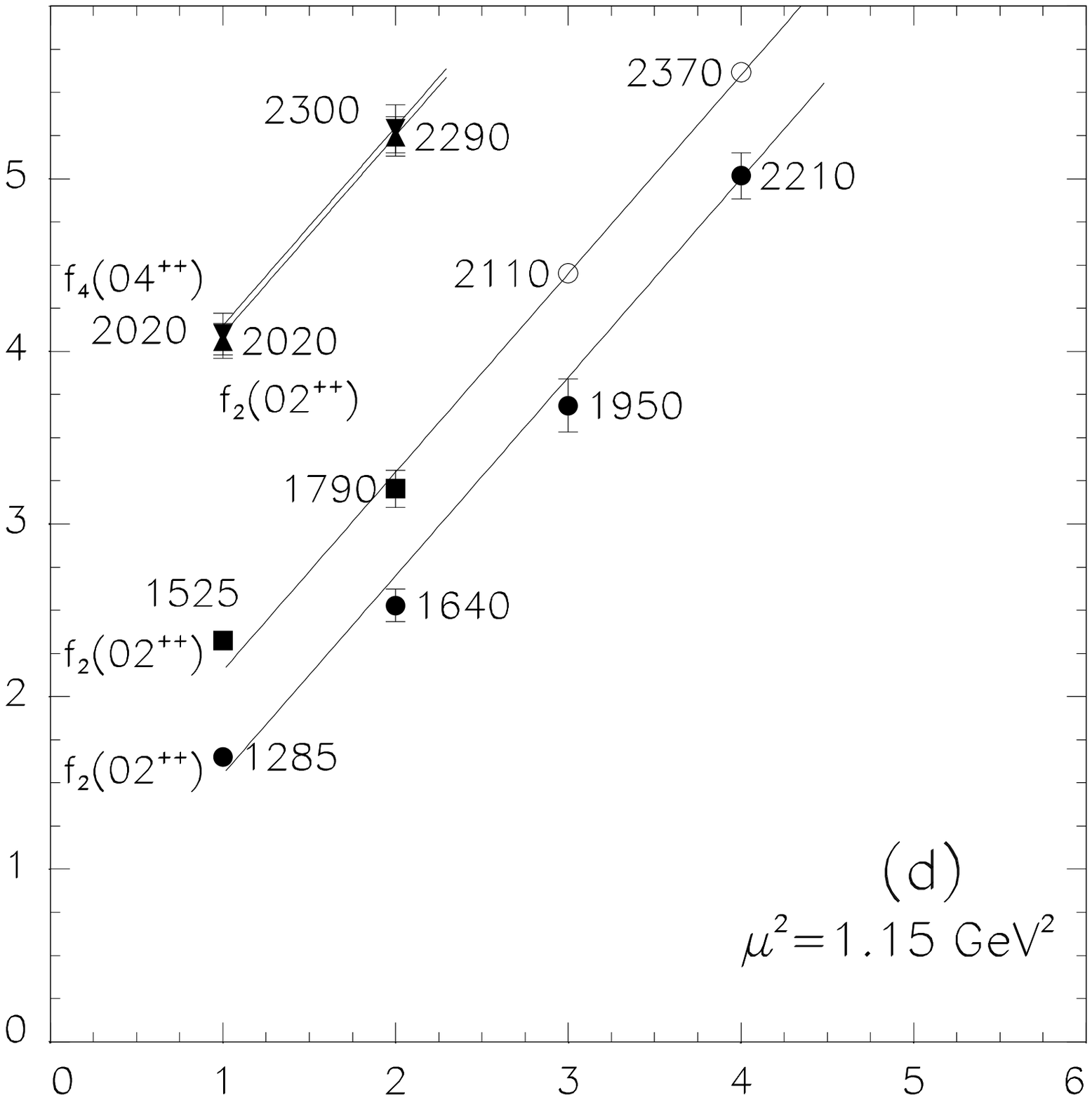,width=8cm}}
\vspace{-1.5cm}
\centerline{\epsfig{file=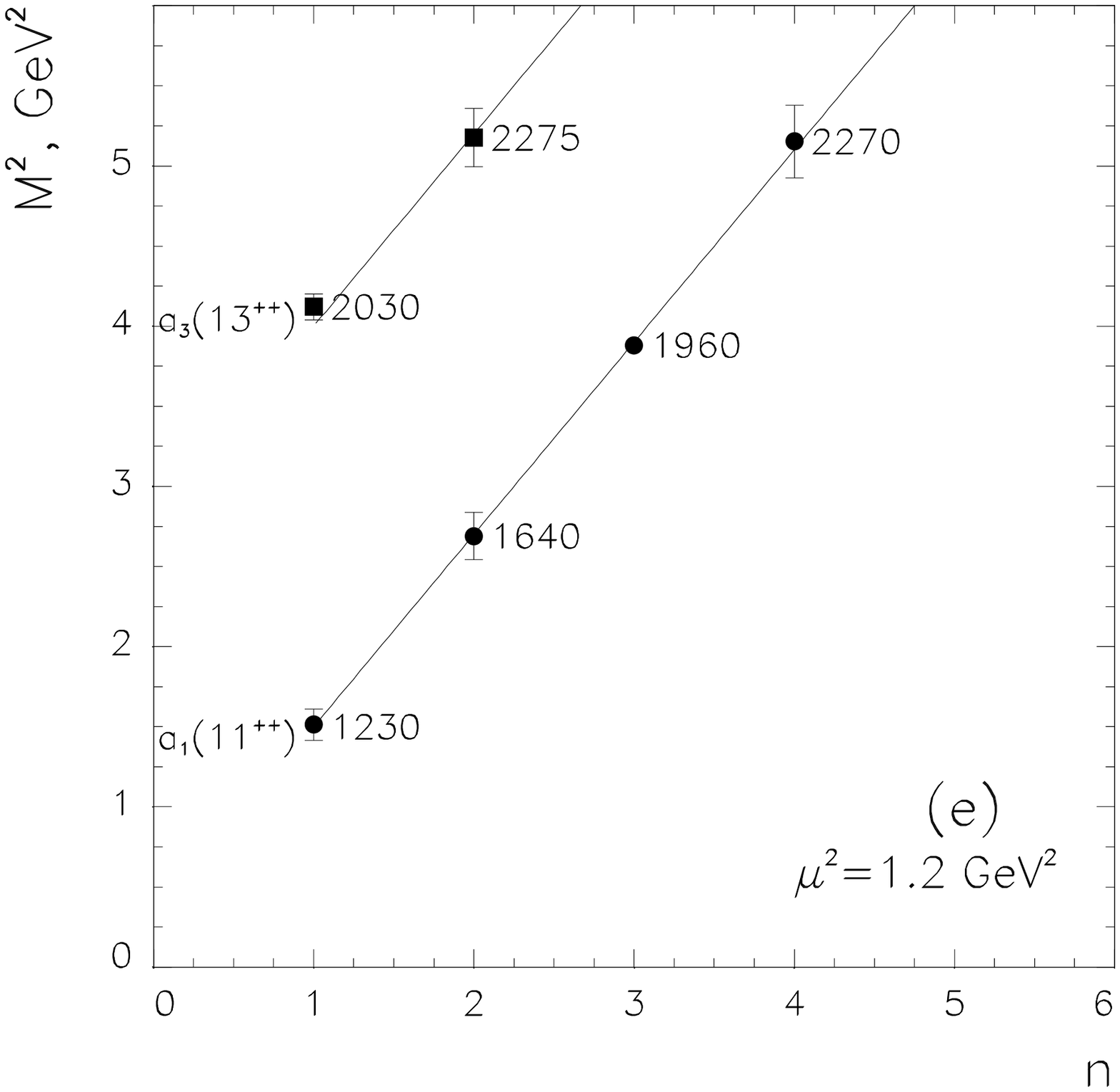,width=8cm}\hspace{-1.5cm}
            \epsfig{file=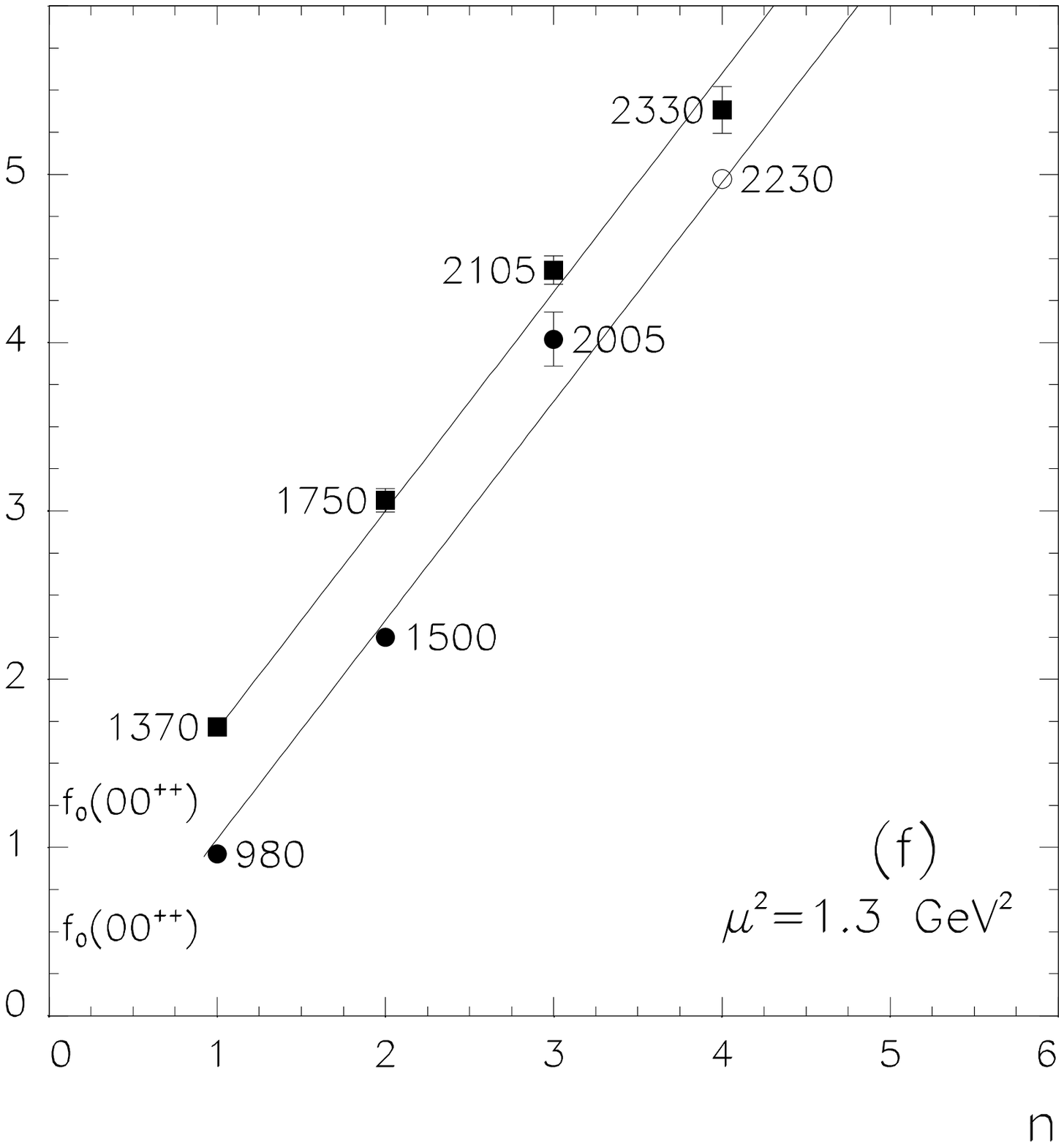,width=8cm}}
\caption{Trajectories of the $(C=-)$-states on the $(n, M^2)$ plane.}
\end{figure}

\newpage
\begin{figure}
\centerline{\epsfig{file=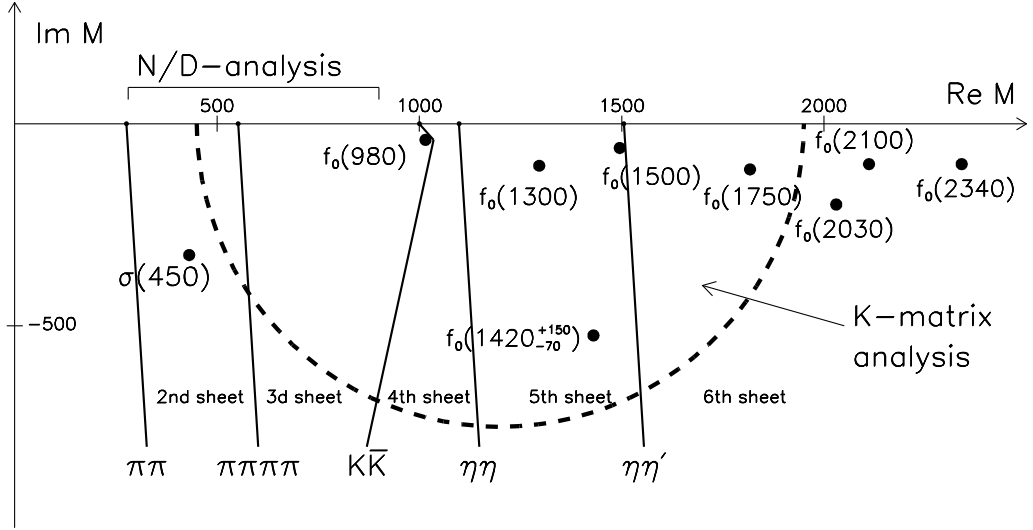,width=14cm}}
\caption{ Complex $M$-plane in the $(IJ^{PC}=00^{++})$ sector. Dashed
line encircle the part of the plane where the $K$-matrix analysis [8]
reconstructs the analytic $K$-matrix amplitude: in this area
the poles corresponding to resonances
$f_0(980)$, $f_0(1300)$, $f_0(1500)$,
$f_0(1750)$ and the broad state $f_0(1420\; ^{+150}_{-70})$ are
located. Beyond this area the light $\sigma$-meson is located
(the position of pole found in the $N/D$ method [6] is
shown) as well as resonances $f_0(2030),f_0(2100),f_0(2340)$ [2].
Solid lines stand for the cuts related to the thresholds
$\pi\pi,\pi\pi\pi\pi,K\bar K,\eta\eta,\eta\eta'$.}
\end{figure}

\newpage
\begin{figure}
\centerline{\epsfig{file=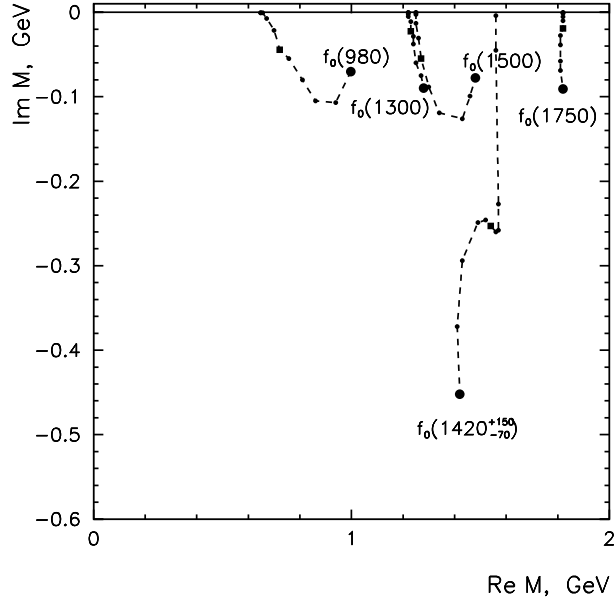,width=9cm}}
\caption{ Complex $M$-plane: trajectories of the poles
for $f_0(980)$, $f_0(1300)$, $f_0(1500)$,
$f_0(1750)$, $f_0(1420\; ^{+150}_{-70})$
during gradual onset of the decay processes.}
\end{figure}

\newpage
\begin{figure}
\centerline{\epsfig{file=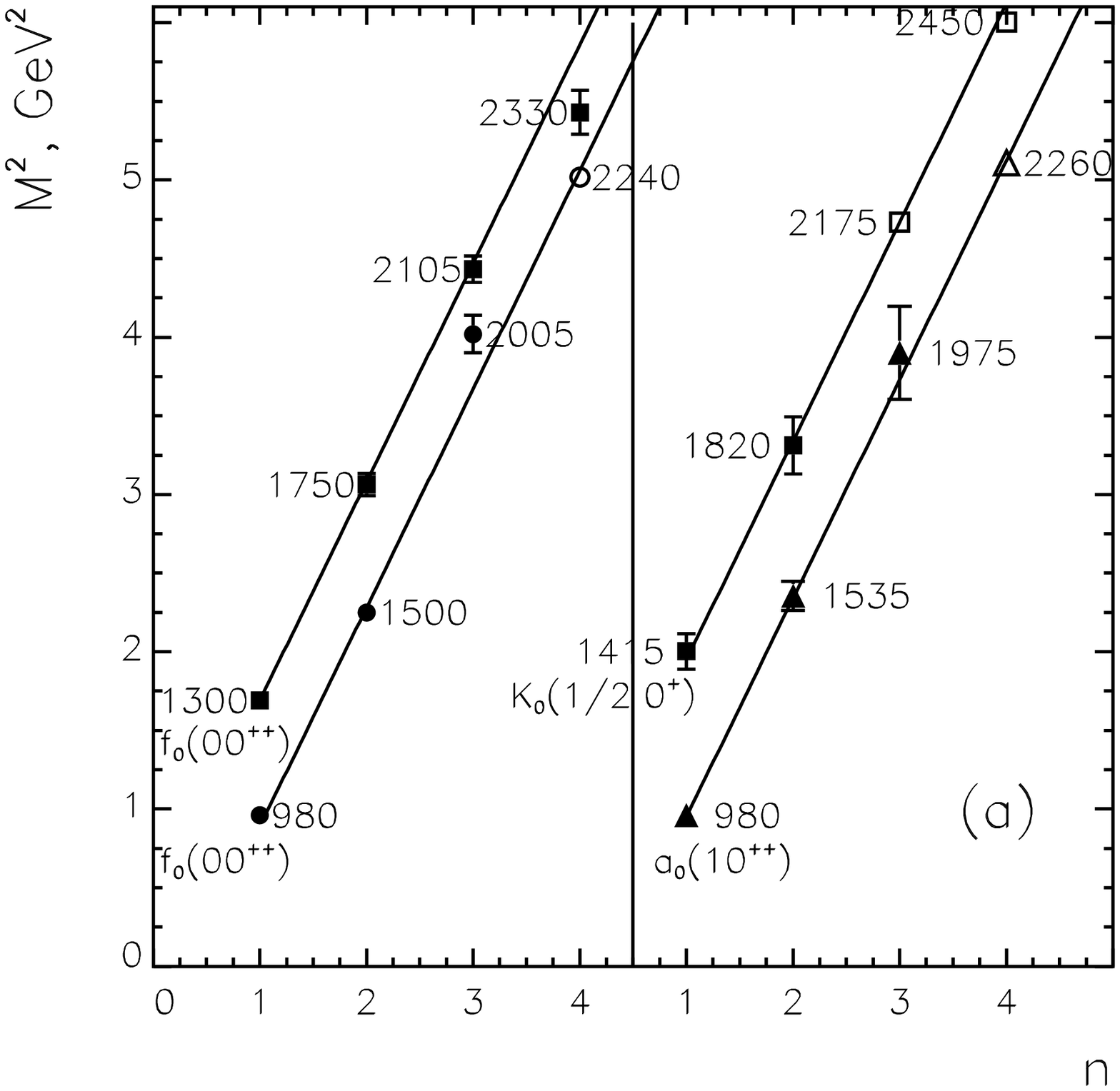,width=8cm}
            \epsfig{file=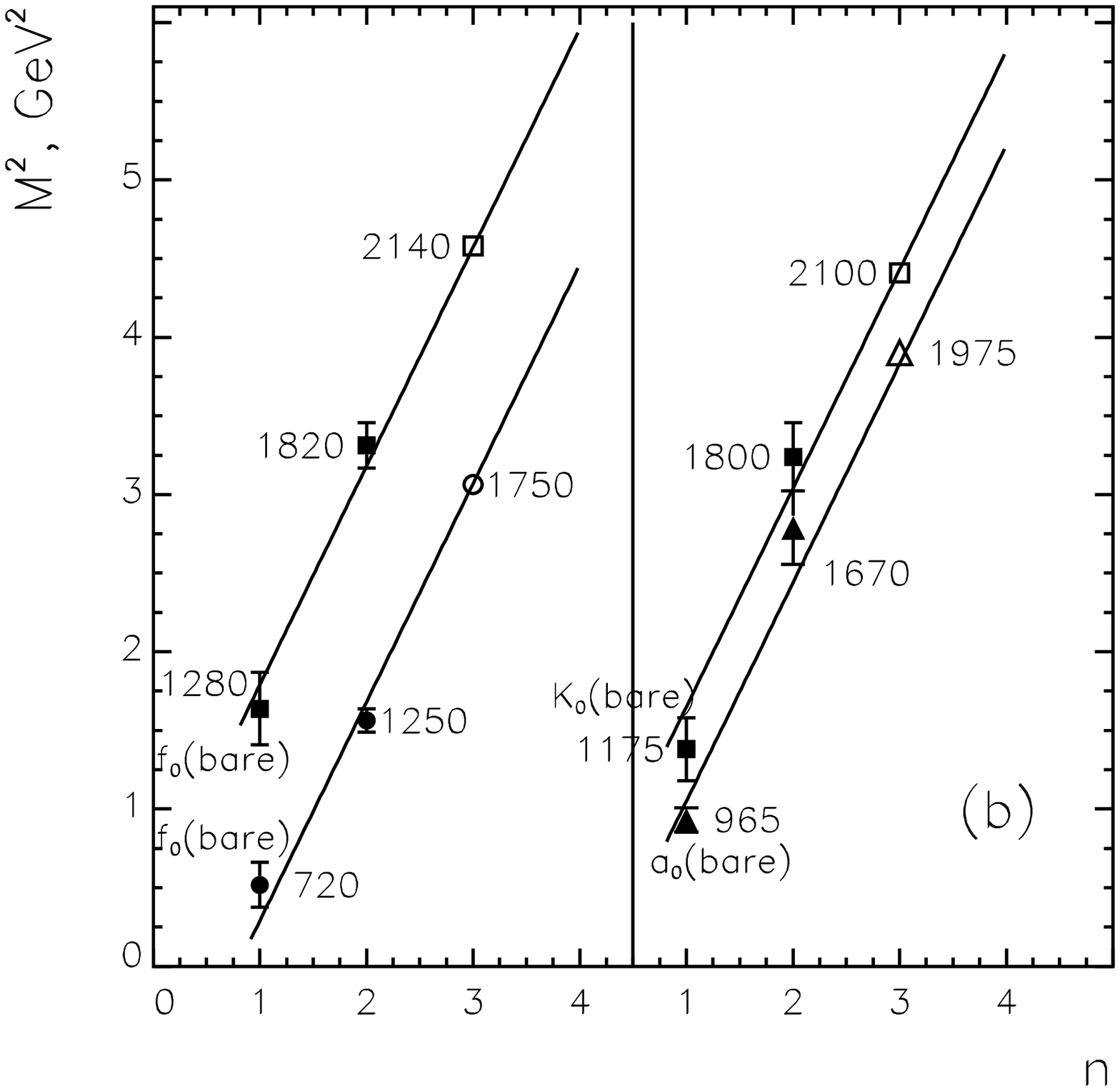,width=8cm}}
\caption{ Linear trajectories in $(n,M^2)$-plane for
scalar resonances (a) and scalar bare states (b).
Open points stand for predicted states.}
\end{figure}

\begin{figure}
\centerline{\epsfig{file=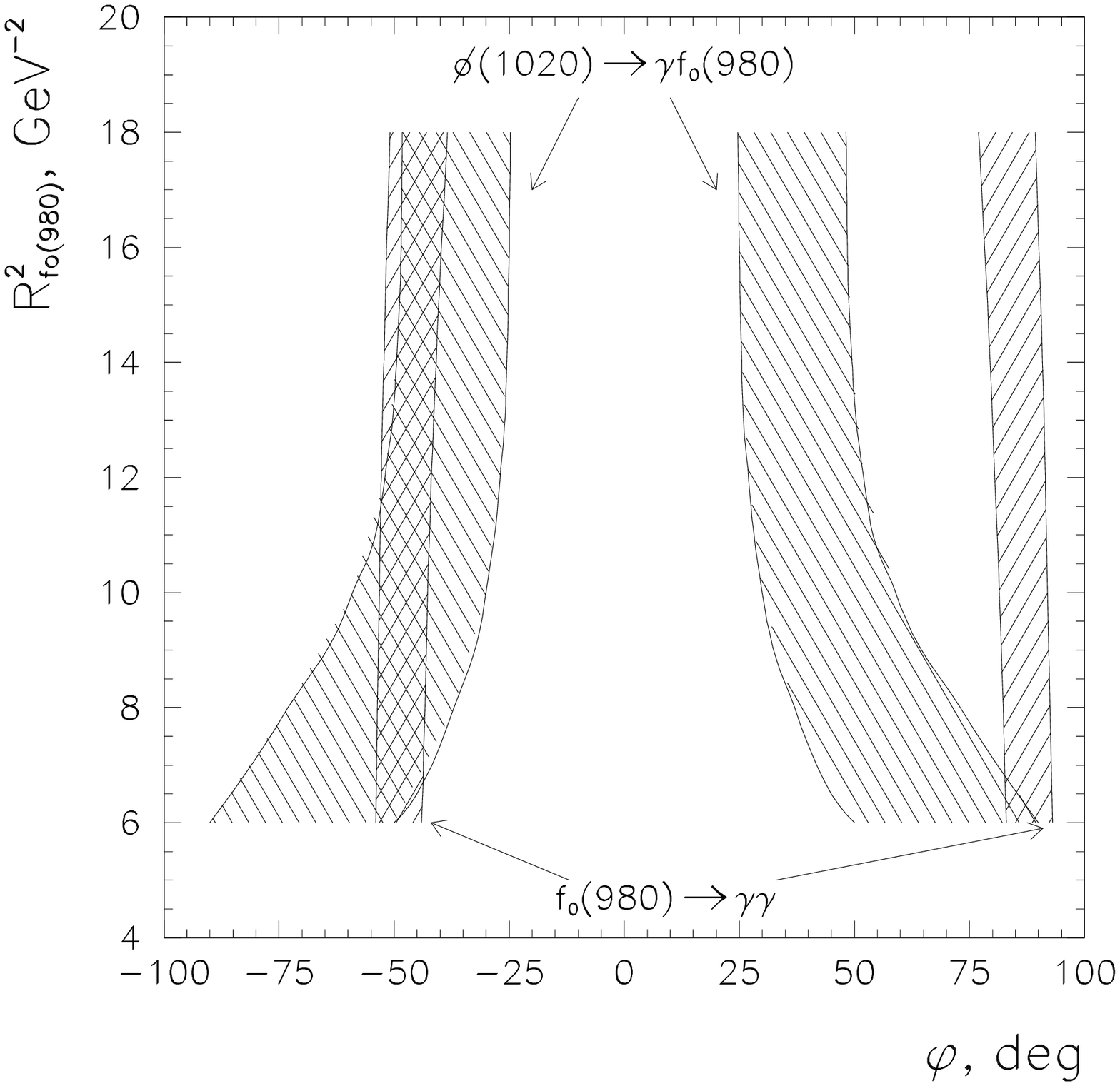,width=8cm}
            \epsfig{file=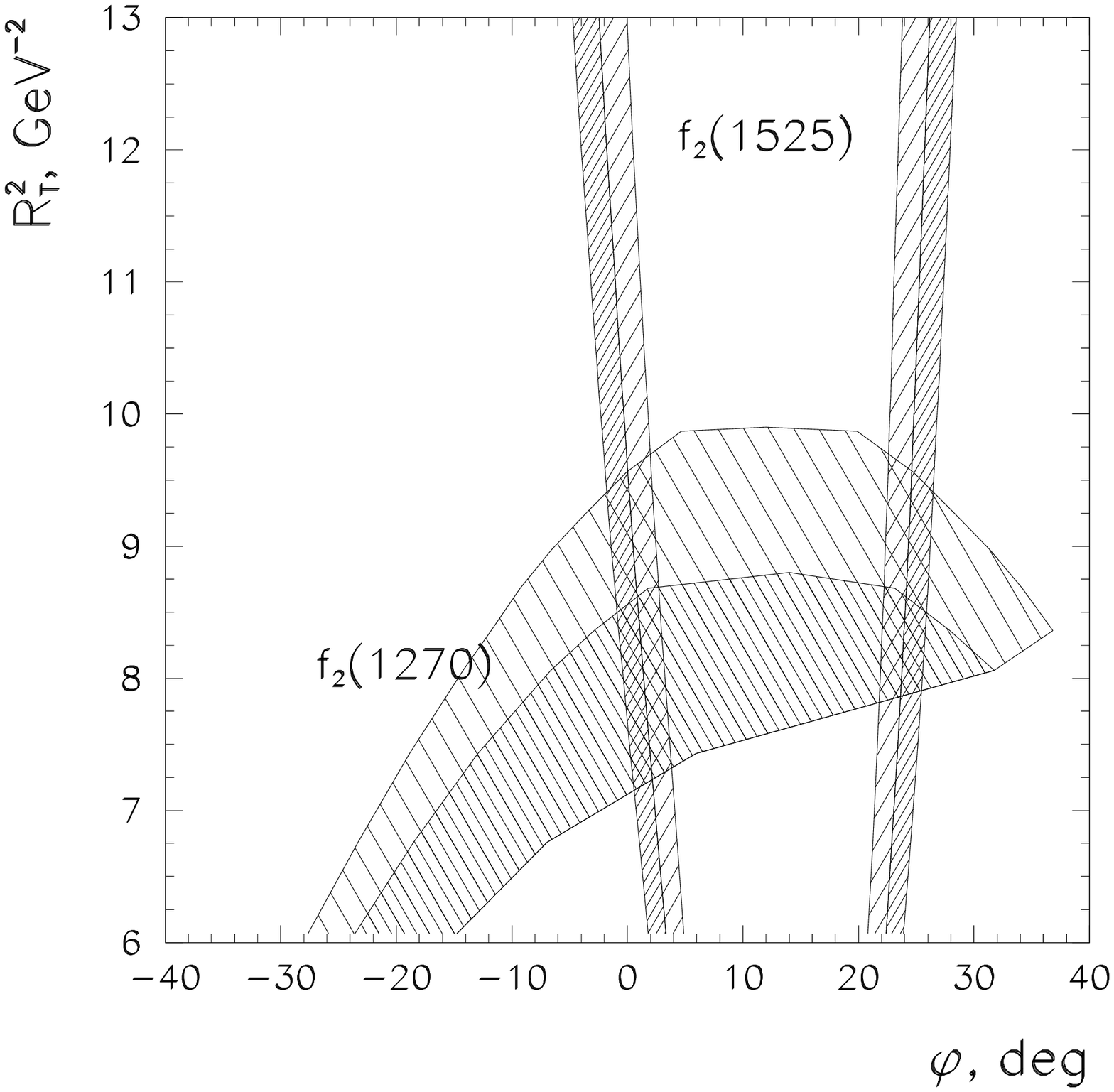,width=8cm}}
\caption{a) The $(\varphi,R^2_{f_0(980)})$-plot: the shaded areas are
the allowed ones for the reactions
$\phi(1020)\to\gamma f_0(980)$ and $ f_0(980) \to \gamma\gamma$.
b) The $(\varphi_T,R^2_T)$-plot for the reactions
$f_2(1270)\to \gamma\gamma$ and $f_2(1525)\to \gamma\gamma$; the
allowed areas are shaded.}
\end{figure}

\end{document}